\documentclass[doublecol]{epl2} 
\newcommand {\bnabla} {\mbox{\boldmath$\nabla$}}

\newcommand {\bcalE} {\mbox{\boldmath$\cal E$}}

\title{Continuum theories of structured dielectrics}
\author{Ralf Blossey \inst{1} \and Rudolf Podgornik \inst{2,3,4}}
\shortauthor{R. Blossey \etal}

\institute{                    
  \inst{1} University of Lille, Unit\'e de Glycobiologie Structurale et Fonctionnelle (UGSF) CNRS UMR8576, 59000 Lille, France\\
  \inst{2} School of Physical Sciences and Kavli Institute for Theoretical Sciences, University of Chinese Academy of Sciences, Beijing 100049, China \\
  \inst{3} CAS Key Laboratory of Soft Matter Physics, Institute of Physics, Chinese Academy of Sciences, Beijing 100190, China  \\
  \inst{4}  Wenzhou Institute of the University of Chinese Academy of Sciences, Wenzhou, Zhejiang 325000, China
}
\abstract{
Aequous dielectrics are ubiquitous in soft- and bio-nano matter systems. The theoretical description of such systems in terms of continuum (`macroscopic')
theory remains a serious challenge. In this perspective we first review the existing
continuum phenomenological approaches that have been developed in the
past decades. In order to describe a path to advance continuum theory beyond these approaches we then take recourse to the Onsager-Dupuis theory of the dielectric behaviour of ice, which, for the case of a solid dielectric, exemplified important
conceptual issues we deem relevant for the development of a more fundamental continuum theory of liquid dielectrics. Subsequently, we discuss our recently
proposed continuum field theory of structured dielectrics, which provides a generalized approach to the dielectric behavior of such systems. }

\begin{document}

\maketitle

\section{Introduction} 

Aqueous dielectrics, especially confined at and/or between macromolecular surfaces and interfaces ~\cite{Gonella2021,Bjorneholm2016, Calero2020,Ruiz2019,Knight2019},  
while being ubiquitous in soft- and bio-matter systems \cite{Holm2001,Dean2014}, still pose a challenge in terms of a consistent macroscopic description. {\sl Grosso modo} the formal difficulties are tied to the dielectric response of water, which is strongly non-local and/or non-linear, and cannot be exhaustively described by a local dielectric approximation. The dielectric properties of aqueous solvent are pertinent to almost every facet of the long-range nanoscale interactions ~\cite{French2010}, since both of the fundamental components, the van der Waals interaction ~\cite{Woods2016} as well as the 
electric double layer interaction ~\cite{Markovich2021}, depend importantly on the static as well as the dynamic dielectric response of the aqueous medium. 

The pursuit of the continuum, macroscopic description of the dielectric properties of water can be roughly partitioned into the phenomenological non-local dielectric function methodology developed extensively by Kornyshev and collaborators \cite{Kornyshev1978,Kornyshev1986,Bopp1996,Bopp1998}, as well as the Landau-Ginzburg-type free energy functionals describing the non-local dielectric response in spatially confined dielectrics  \cite{Kornyshev1997,Maggs2006, Mertz2007}, the two approaches being of course equivalent in their linear limiting forms in the bulk but differing in confined, inhomogeneous systems.    

In this short perspective we first briefly review these approaches and some of their recent
applications. Subsequently, we argue that a continuum field theory approach to
structured liquid dielectrics can be designed by taking recourse to an exemplary
theory for the dielectric properties of solid ice, the Onsager-Dupuis theory 
\cite{Onsager1960,Onsager1962}. Following a concise description of this theory of the solid phase of water we then present the key essentials of our recent formulation of a general field theory of liquid aqueous dielectrics. In particular we stress the fundamental similarity of both non-local and non-linear theories.

\section{Phenomenological theories}  

Phenomenological functionals of polarization have featured prominently in the development of the electrostatic theory of structured dielectrics starting from the seminal contributions of Marcus in the 1950's on electron transfer theory based upon the local dielectric response theory \cite{Marcus1956,Marcus1956-2}, also discussed in detail by Felderhof \cite{Felderhof1977}. These theories describe polarization on the basis of harmonic functionals in which dielectric properties are local. However, on the nanometer
scales relevant to bio-nano systems, we want to address the dielectric response which is non-local and needs to be described in terms of a non-local constitutive relation for the dielectric displacement field
\begin{equation}
    {\bf D}({\bf r}) = \int d^3{\bf r'} \varepsilon({\bf r},{\bf r'}) {\bf E}({\bf r'}),
\end{equation}
expressed as a space-integral over the two-argument dielectric function $\varepsilon ({\bf r},{\bf r'})$, or, in Fourier-space ${\bf D}(q) = \varepsilon(q){\bf E}(q)$, 
which makes the scale-dependent dielectric behaviour even more transparent. The
function $\varepsilon(q)$ typically drops from the water bulk value $\varepsilon \sim 80 $ to a value of about 20 at the Bjerrum length of a monovalent ion \cite{Maggs2006}, while at even smaller length-scales it diverges and becomes negative, so that 
a first level of treatment of non-local dielectric properties resides in the
approximations made for $\varepsilon(q)$ \cite{Bopp1998,Hildebrandt2004,Schaaf2016}.

 Maggs and Everaers \cite{Maggs2006} showed that the non-local approach can be equivalently formulated in terms of free energy functionals of the type
\begin{eqnarray}
U = \frac{1}{2}\int d^3{\bf r}({\bf D}({\bf r}) - {\bf P}({\bf r}))^2 + \nonumber \\
\frac{1}{2}\int d^3{\bf r} d^3{\bf r'}{\bf P}({\bf r}) K({\bf r},{\bf r'}){\bf P}({\bf r'})
\end{eqnarray}
where the integral kernel $K$ relates to $\varepsilon$ via 
\begin{equation}
    \varepsilon_{ij}(q) = \delta_{ij} + K_{ij}^{-1}(q)\, 
\end{equation}
with the standard decomposition
\begin{eqnarray}
K_{ij}(q) = K_{\parallel}(q) \frac{q_i q_j}{q^2} + K_{\perp}(q) \left( \delta_{ij} - \frac{q_i q_j}{q^2} \right)
\end{eqnarray}
in Fourier space. Looking at the stationary points of functional Eq. (2) under the
constraint of Gauss' law, these  authors could show that the functional equivalent to the Marcus functional can be derived as
\begin{eqnarray} \label{functional}
    U_p = \frac{1}{2} \int d^3{\bf r}d^3{\bf r'} \frac{\bnabla {\bf P}({\bf r}) \cdot \bnabla {\bf P}({\bf r'})}{|{\bf r} - {\bf r'}|} + \nonumber \\
    \frac{1}{2}\int d^3{\bf r} d^3{\bf r'}{\bf P}({\bf r}) K({\bf r},{\bf r'}){\bf P}({\bf r'}) + U({\bf P},{\bf E_0})  
\end{eqnarray}
where the last term contains contributions from the bare electric field 
${\bf E_0}$ generated by the free charges in vaccuum.  

Eq. (\ref{functional}) can now serve as the basis for a variety of phenomenological models via appropriate choices of the kernel $K_{ij}$ in terms of a Ginzburg-Landau expansion.
A general choice of the harmonic term with kernel $K_{ij}$ is provided by the expression
\begin{eqnarray} \label{maggs}
U_K &\equiv& \frac{1}{2}\int d^3{\bf r}
\left[K {\bf P}^2({\bf r}) + \kappa_l(\bnabla \cdot {\bf P}({\bf r}))^2 \right. \nonumber \\
& + & \left.\kappa_c (\bnabla \times {\bf P}({\bf r}))^2 + 
\alpha(\bnabla(\bnabla \cdot {\bf P}({\bf r})))^2\right] 
\end{eqnarray}
For this model this leads to longitudinal and transverse dielectric susceptibilities 
$\chi_{\parallel}(q)$, $\chi_{\perp(q)}$ in Fourier space of the form
\begin{equation}
    \chi_{\parallel}(q) \equiv 1 - \frac{1}{\varepsilon_{\parallel}(q)} 
    = \frac{1}{1 + K + \kappa_lq^2 + \alpha q^4}
\end{equation}
and
\begin{equation}
    \chi_{\perp}(q) \equiv \varepsilon_{\perp}(q) - 1 = \frac{1}{K + \kappa_c q^2}\, ,
\end{equation}
where $\varepsilon_{\parallel}(q)$ and $\varepsilon_{\perp}(q)$ are the corresponding longitudinal and
transverse dielectric functions.
Model (\ref{maggs}) has been used by Berthoumieux and Maggs to discuss
fluctuation-induced structural interactions \cite{Berthoumieux2015}.

Further examples of functionals defined in this vein have recently been discussed by
Berthoumieux and collaborators. This series of works deals with the formulation of a Gaussian water model based on two order parameter description in terms of fluid density and polarization \cite{Berthoumieux2018} following the early work by \cite{Kornyshev1997}. A non-local and nonlinear theory of water solvation of ions was  discussed in \cite{Berthoumieux2019}. The Landau-Ginzburg continuum approach was also validated against explicit molecular simulations \cite{Vatin2021} and applied to water confined between surfaces at nanoscale separations \cite{Monet2021}, by complementing the non-local Landau-Ginzburg continuum approach in the bulk with local surface terms,
very much in the philosophy of the general Landau-Ginzburg theories of surface phase transitions. Subsequent work considered the effect of polarization saturation which is required if one wants to properly describe high excitations fields \cite{Berthoumieux2021}. Nonlinear polarization functionals can
be derived if one starts from the dual formulation of soft matter electrostatics 
in which one introduces the polarization field via a Legendre transform starting from the Poisson-Boltzmann theory \cite{Maggs2012,Blossey2018}.

While these results are already promising in terms of applications of the phenomenological approach to different physical situations, these essentially symmetry-based models still lack the key feature of a systematic derivation. Before we discuss our recent proposal of how this can be achieved in the case of liquid dielectrics, and thus define a path to formulate very general continuum models containing all physically relevant effects, we first make a little detour and turn to the Onsager-Dupuis theory of the dielectric properties of crystalline ice.

\section{Onsager-Dupuis theory of dielectric properties of ice}

\begin{figure}
\onefigure[width= 8cm]{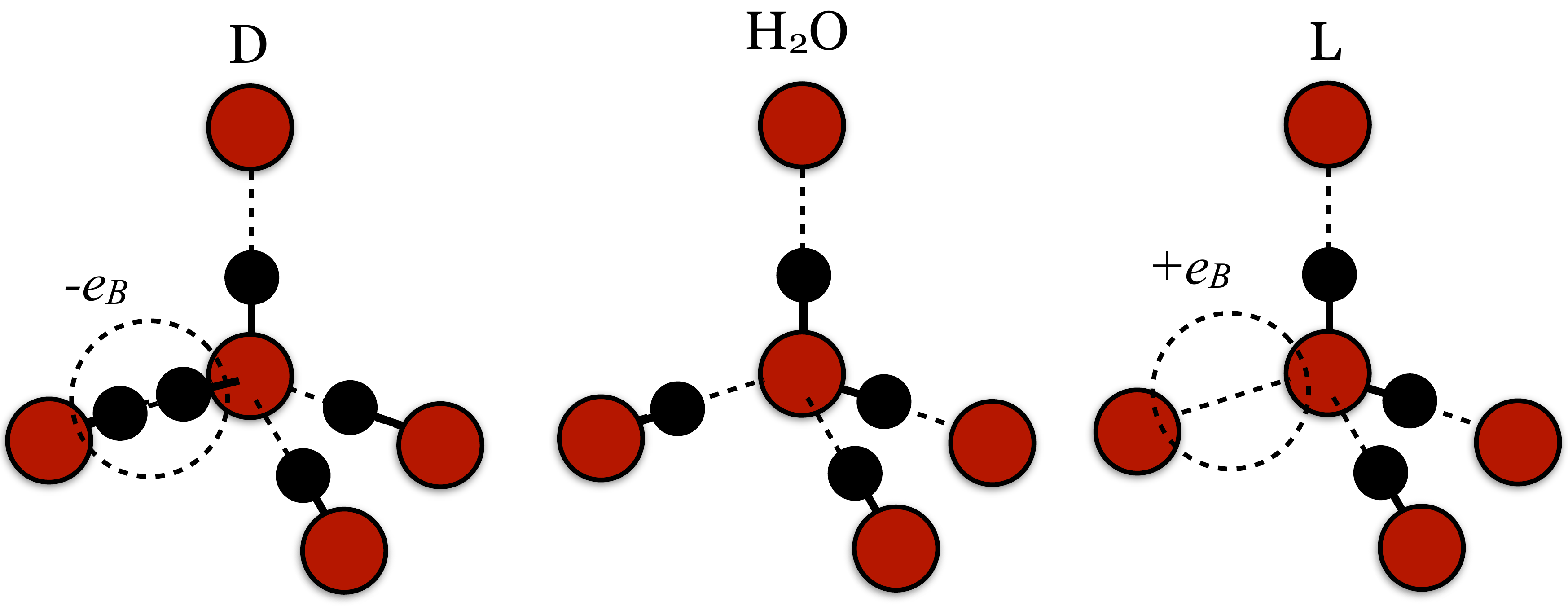}
\caption{D and L Bjerrum defects, with two (doppelt) or no (leer) protons between neighboring water molecules, as well as the Bernal-Fowler fully bonded water molecule, in a hexagonal ice. One molecules and its first four neighbors are shown. The Bjerrum defects are then sources and sinks of configurational polarization, Eq. (\ref{Bjerrum1}), with polarization charge of $\pm e_B$. }
\label{fig1}
\end{figure}

In crystalline ice at a finite temperature the polarization and electrostatic fields are coupled by the structural interaction mediated by the free and salt-ion bound Bjerrum structural defects in the ideal hydrogen bonded lattice of ice, see Fig. \ref{fig1}. 
A mean-field statistical mechanical formulation of the dielectric properties of ice was proposed by Onsager and Dupuis \cite{Onsager1960,Onsager1962} and later further elaborated and generalized by Gruen and Mar\v celja \cite{Gruen1983, Gruen1983-2}. 

The total polarization can be decoupled into its atomic (${\bf P}_a = \varepsilon_0 (\varepsilon_{\infty} -1 ) {\bf E}$) and configurational (${\bf P}_c$) contributions, where ${\bf E}$ is the electric field vector. The two, in distinction to the phenomenological theories, therefore represent {\sl independent degrees of freedom}. 
The origin of $\varepsilon_{\infty}$ is the atomic polarization of the water molecules. The total polarization is then given by 
\begin{equation}
    {\bf P} = \varepsilon_0(\varepsilon_{\infty}-1) {\bf E} + {\bf P}_c.
\end{equation}
In addition one also has the standard Poisson equation in the form  
\begin{eqnarray}
\bnabla\cdot{\bf D} = e_0 N \left(c_+ - c_-\right),
\label{Poisson1}
\end{eqnarray}
with ${\bf D} = \varepsilon\varepsilon_{\infty} {\bf E} + {\bf P}_c$, where $c_+, c_-$ are the relative concentrations of the ice-dissolved cations and anions, while $N$ is the concentration of the water moelcules in ice. 

One furthermore notes that in an ideal ice according to the Bernal-Fowler rules, stipulating that each water molecule is neutral and fully hydrogen bonded to its four nearest neighbors, one has
$\bnabla\cdot{\bf P}_c = 0$. However, at finite temperatures the D (doppelt) and L (leer) structural Bjerrum defects, see Fig. \ref{fig1}, present the sources of configurational polarization, leading to \cite{Onsager1960,Onsager1962}
\begin{eqnarray}
\bnabla\cdot{\bf P}_c = e_B N \left(c_L - c_D + n (c_+ - c_-)\right),
\label{Bjerrum1}
\end{eqnarray}
where $e_B$ is the ``polarization charge" given by the dipolar moment of the water molecule and the oxygen-oxygen distance in the ice lattice. Onsager and Dupuis calculate $e_B = \sqrt3 \mu/d$, where $d$ is the distance between water molecules and $\mu$ is the dipole moment, leading to $e_B \simeq 0.35 e_0$, where $e_0$ is the electronic charge \cite{Nagle1974}.  
The ``hydration number" $n$ is the number of Bjerrum defects associated with dissolved anions and cations, assumed to have identical hydration shells, and $c_L, c_D$ are relative concentrations of the Bjerrum defects. In what follows we will simplify the matter by considering the ``hydration number" as zero, {\sl i.e.} $n = 0$, pertaining to ions without any hydration shell.

The interactions between all the effective charges, the dissolved ions as well as the Bjerrum defects, are assumed to be of the Coulomb form
\begin{equation}
u({\bf x} -{\bf x}') = ({1}/{4\pi {\textstyle \varepsilon_{\infty}\varepsilon_0})~ \vert {\bf x} -{\bf x}' \vert}^{-1}\,,
\label{Coulomb}
\end{equation}
but with the high frequency dielectric constant. 

Evaluating the partition function of ice with local non-vanishing polarization and structural defects on the mean-field level (equivalent to neglecting the formation of rings of water molecule in the crystal - dendritic lattice),  one ends up with the free energy density in the form \cite{Gruen1983,Gruen1983-2} 
%\begin{widetext}
\begin{eqnarray}
{\cal F} &=& {\textstyle\frac12} \varepsilon_0 \varepsilon_{\infty} E^2 - k_BT \log\frac{W(c_D, c_L, c_+, c_-)}{W_0}  \nonumber \\ 
&& + \sum_{i = D,L,+,-} \mu_i N (c_i - c_{i0})
\label{totalF}
\end{eqnarray}
%\end{widetext}
where $W(c_D, c_L, c_+, c_-)$ is the number of configurations of polarized ice with  a given configuration of D,L defects and cations, anions, $W_0$ is the number of configurations of ideal ice and $\mu_i$ are the chemical potentials of the 
$i = D,L,+,-$ species (D and L Bjerrum defects, ice dissolved cations and anions). Index zero in concentrations refers to the reservoir assuming $c_{D0} = c_{L0}$ and $c_{+0 }= c_{-0} = c_{I0}$. Note also that the electrostatic part of the free energy, the first term in Eq. (\ref{totalF}), is given by the Coulomb interaction screened with the high-frequency dielectric constant.

Assuming furthermore a homogeneous crystal and {by minimizing} the above free energy w.r.t. the external field, one derives the dielectric constant of the Slater-Takagi-Bethe form \cite{Nagle1974}
\begin{equation}
\varepsilon_0 (\varepsilon - \varepsilon_{\infty}) =  \frac{N \mu^2}{k_BT} \frac{1-  c_{D0}}{1 +  c_{D0}},
\end{equation}
where $\varepsilon$ is the full dielectric constant with the ice configurational component included. 
Minimizing now the total free energy then yields two equations for the electrostatic and the polarization potentials 
\begin{eqnarray}
{\bf E} = - \bnabla \phi \quad {\rm and} \quad {\bf P}_c = \varepsilon_0(\varepsilon-\varepsilon_{\infty}) \bnabla \phi_c
\label{potentials}
\end{eqnarray}
that can be rewritten as
\begin{eqnarray}
\bnabla \cdot {\bf P}_c &=& \frac{\partial}{\partial \phi_c} p(\phi, \phi_c) \nonumber\\
\varepsilon_0  \varepsilon_{\infty} \bnabla \cdot {\bf E}  &=& - \frac{\partial}{\partial \phi} p(\phi, \phi_c),
\label{BjerrumPB}
\end{eqnarray}
where $p(\phi, \phi_c)$ is the osmotic pressure of the ice which in the Onsager-Dupuis theory is given by the van't Hoff expression in terms of the ion and Bjerrum defect density
\begin{eqnarray}
p(\phi, \phi_c) &=& 2 k_BT~ n_{D0} \cosh{\beta{e_B} (\phi + \phi_c)} + \nonumber\\
&& ~~~~~+ 2 k_BT~\cosh{\beta {e_0} \phi}.
\label{vtH1}
\end{eqnarray}
It is possible to show that this  osmotic pressure is a limiting form of a more general expression equivalent to a lattice gas pressure, if one considers explicitly the finite number of crystalline sites 
\cite{Podgornik1986}. 
The Euler-Lagrange equations, Eqs. (\ref{BjerrumPB}), can be rewritten in the form of two coupled Poisson-Boltzmann-type equations
%\begin{widetext}
\begin{eqnarray}
\varepsilon_0 (\varepsilon - \varepsilon_{\infty}) \bnabla^2 \phi_c &=& 2 n_{D0} e_B \sinh{\beta{e_B} (\phi + \phi_c)} \nonumber\\
\varepsilon_0 \varepsilon_{\infty} \bnabla^2 \phi &=& 2 n_{I0} e_0 \sinh{\beta{e_0} \phi} + \\
&& + 2 n_{D0} e_B \sinh{\beta{e_B}  (\phi + \phi_c)}, \nonumber 
\label{PBEQUS}
\end{eqnarray}
%\end{widetext}
first derived in this specific form by Babcock and Longini \cite{Babcock1972}. Above $n_{I0} = N c_{I0}$ and $n_{D0} = N c_{D0}$ are the bulk concentrations of the Bjerrum defects and salt ions. Linearizing the non-linear system of Poisson-Boltzmann equations Eq. (\ref{PBEQUS}) one obtains a system of two coupled linear equations for the conjugated potentials 
\begin{eqnarray}
\bnabla^2 \phi_c &=&  \xi_B^{-2} (\phi + \phi_c), \nonumber\\
\bnabla^2 \phi &=& {\textstyle\frac{\varepsilon}{\varepsilon_{\infty}}} \lambda_D^{-2} \phi  + \left( {\textstyle\frac{\varepsilon}{\varepsilon_{\infty}}} - 1 \right)\xi_B^{-2} (\phi + \phi_c), 
\end{eqnarray}
derived first by Onsager and Dupuis by a different route \cite{Onsager1960,Onsager1962}, and also rederived latter \cite{Manciu2004,Manciu2005,Paillusson2010}.
The above two equations clearly attest to the two independent degrees of freedom: the  configurational polarization and the electric field.

Obviously one can now introduce two screening lengths: the Debye length $\lambda_D$ and the structural length $\xi_B$ as
\begin{eqnarray}
\lambda_D^{-2} = \frac{2 \beta e_0^2~ n_{I0}}{\varepsilon_0\varepsilon} \quad {\rm and} \quad \xi_B^{-2} = \frac{2 \beta e_B^2~ n_{D0}}{\varepsilon_0 (\varepsilon-\varepsilon_{\infty})}
\end{eqnarray}
and the solution of the linearized equations then 
leads to exponentially decaying solutions with the characteristic exponents obtained from 
\begin{eqnarray}
\lambda^4 - \lambda^2  \Big( \lambda_D^{2} + \frac{\varepsilon_{\infty}}{\varepsilon} \xi_B^2\Big) +  \left({\xi_B}{\lambda_D}\right)^2 = 0,
\label{charexpo}
\end{eqnarray}
implying in addition a non-local dielectric function of the form
\begin{eqnarray}
\varepsilon(k) = \varepsilon_{\infty} + \frac{\varepsilon - \varepsilon_{\infty}}{1 + \frac{\varepsilon}{\varepsilon_{\infty}} (\xi_B k)^2} +  \frac{\varepsilon}{(\lambda_D k)^2},
\end{eqnarray}
equivalent to a linear superposition of the Inkson non-local (static)  dielectric response of the Lorentzian type for the structural component \cite{Inkson1972}  
and the Debye electrostatic screening of ionic charges  \cite{Kornyshev1978}.

The Onsager-Duipuis theory therefore describes the electrostatic interactions between ions and the structural interactions between mobile and ion-associated Bjerrum defects in ice, {\sl via} an effective electrostatic interaction between Bjerrum polarization charges.
Obviously the Onsager-Dupuis theory embodies not only non-local dielectric effects but also the non-linear dielectric response of a system. It leads to {\sl two coupled equations} for the electrostatic field and the dielectric polarization field, and in general cannot be reduced to either the non-local dielectric function methodology or the local Landau-Ginzburg polarization functional. 

\section{A field theory of structured liquid dielectrics}

In liquid dielectrics one cannot define structural defects because the liquid is disordered. The consideration of its statistical and dielectric properties thus necessarily starts from a different vantage point. This has been first consistently formulated in Ref. \cite{Blossey2022} by assuming a mixture of dipolar solvent molecules and monopolar electrolyte ions, interacting via electrostatic and non-electrostatic structural interatctions.

\begin{figure}
\onefigure[width= 8cm]{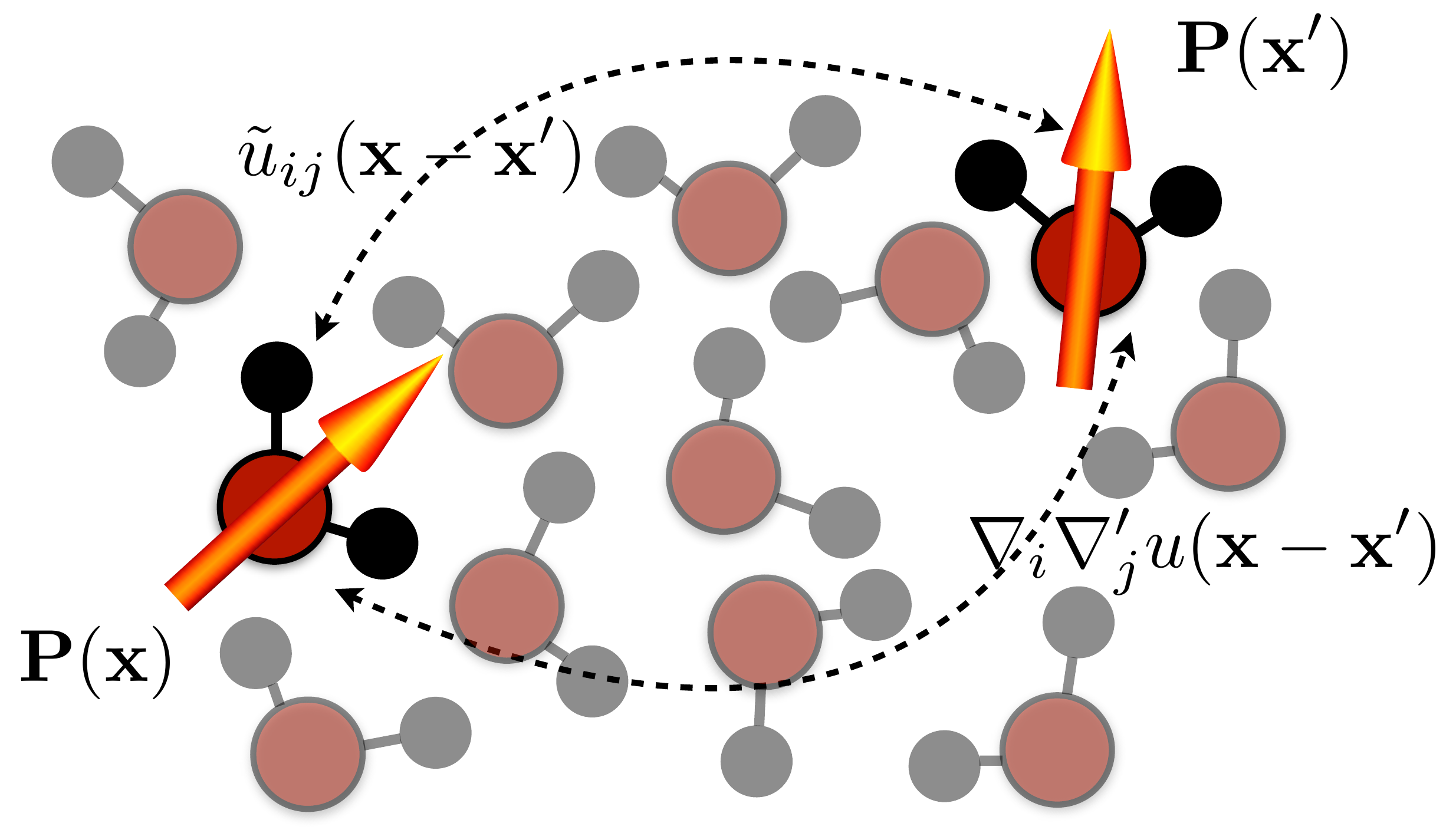}
\caption{Interaction between two solvent molecule dipoles (cylindrical arrows) in a structured dielectric: Coulomb dipolar interaction ($\bnabla_i\bnabla_j'u({\bf x} - {\bf x}')$) and pure non-local structural interaction ($\tilde{u}_{ij}({\bf x} - {\bf x}')$), Eqs (\ref{hamilt0}), (\ref{hamilt1}).}
\label{fig2}
\end{figure}

An aqueous electrolyte solution is first of all described by two independent ``order parameters",  corresponding to two independent degrees of freedom, the polarization 
\begin{eqnarray}
{\bf P} = {\bf D} - \varepsilon_0 {\bf E}, 
\end{eqnarray}
and the total charge density
\begin{equation}
\bnabla\cdot{\bf D} = e_0 (n_+ - n_-) + \bnabla\!\cdot{\cal \bf P},
 \label{cole3}
\end{equation}
where $n_+,n_-$ are the densities of the univalent ions, and ${\bf D}, {\bf E}$ are the standard dielectric displacement and electric field vectors. Instead of the Coulomb interactions between the ions and the structural defects, we now have the non-local  structural interactions 
\begin{equation} {\textstyle{\textstyle\frac12}} \int\!\!\int_V d{\bf x} d{\bf x}' ~ {\cal \bf P}_{i}({\bf x})\tilde u_{ij}({\bf x} -{\bf x}') {\cal \bf P}_{j}({\bf x}'), 
\label{hamilt0}
\end{equation}
where we assumed that the tensorial part $\tilde u_{ij}({\bf x} -{\bf x}') $ is a {\it short-range, non-electrostatic potential}, and the standard  electrostatic interaction,  given by the standard Coulomb form
\begin{equation} {\textstyle{\textstyle\frac12}} \int\!\!\int_V d{\bf x} d{\bf x}' ~ \bnabla\cdot{\bf D}({\bf x})~u({\bf x}\!-\!{\bf x}')~ \bnabla'\cdot{\bf D}({\bf x}').
\label{hamilt1}
\end{equation}
Eqs. (\ref{hamilt0}), (\ref{hamilt1}) imply a Coulomb dipolar interaction ($\bnabla_i\bnabla_j'u({\bf x} - {\bf x}')$) and pure non-local structural interaction ($\tilde{u}_{ij}({\bf x} - {\bf x}')$). 
Here, we take the Coulomb interaction with the non-configurational high-frequency dielectric constant,  $\varepsilon_{\infty}$ as in the Onsager-Dupuis model, {\sl i.e.}
\begin{equation}
u({\bf x} -{\bf x}') = ({1}/{4\pi {\textstyle \varepsilon_{\infty}\varepsilon_0})~ \vert {\bf x} -{\bf x}' \vert}^{-1}\,.
\label{Coulomb}
\end{equation}
The free energy density on the mean-field level is then obtained as
\begin{eqnarray}
{\cal F} &=& - {\textstyle\frac12} \int {\cal  E}_{i}({\bf x})\tilde u_{ij}({\bf x} -{\bf x}') {\cal E}_{j}({\bf x}')d^3{{\bf x}'} - \nonumber\\
&& - {\textstyle\frac12} \varepsilon_0 \varepsilon_{\infty} (\bnabla\phi)^2 ~- p(\phi, {\cal E}),
\end{eqnarray}
where $p(\phi, {\cal E})$ is related to the single-particle partition function, while ${\cal E}$ and $\phi$ are auxiliary (Lagrange fields) potentials defining the polarization and the charge density.  Minimizing the total free energy then yields the following two equations: the constitutive relation
\begin{eqnarray}
%{\cal \bf P}({\bf x}) &=& - %\frac{\partial V^*}{\partial u} \frac{\partial u}{\partial {\cal E}^*({\bf x})} 
%(\beta p)^2  \left({\cal E}^*({\bf x})-\bnabla \phi^*({\bf x})\right) \left(\frac{1}{u}\frac{\partial v^*}{\partial u}\right) 
&& {{\cal \bf P}} = - \frac{\partial p(\phi, {\cal E})}{\partial  {\cal E}} 
\end{eqnarray}
and the Poisson equation
\begin{eqnarray}
&& %{\rho}({\bf x}) - \bnabla \cdot {\cal \bf P} ({\bf x}) 
\varepsilon_0 \bnabla\cdot{\bf E} = -  \frac{\partial p(\phi, {\cal E})}{\partial \phi} -  \bnabla \left(\frac{\partial  p(\phi, {\cal E})}{\partial {\cal E} }\right),
\label{nonlineeq}
\end{eqnarray}
where $p(\phi, {\bcalE}) = p({\cal E} + {\bf E},\phi) = p({\cal E} - \bnabla \phi,\phi)$ is now the osmotic pressure of the aqueous solution, composed of the Langevin dipolar and the ion contributions \cite{Abrashkin2007}
\begin{eqnarray}
p(\phi, {\cal E}) 
= \lambda \frac{\sinh{\beta p ~\vert {\cal E} -\bnabla \phi  \vert}}{\beta p ~\vert {\cal E} -\bnabla \phi  \vert}+ 2 \lambda_s  \cosh{\beta e_0 \phi}, %\nonumber \\
\label{Xi-2}
\end{eqnarray}
with $\lambda, \lambda_s$ the absolute activities of the dipolar solvent and the dissolved ions. 
It is related to the van`t Hoff expression, Eq. (\ref{vtH1}), for the osmotic pressure of ions and Bjerrum defects in the Onsager-Dupuis theory. The auxiliary field ${\cal E}$ is defined as 
\begin{eqnarray}
\label{pol11}
{\cal E}({\bf x}) &=&  - \int_V d{\bf x}' \tilde u_{ij}({\bf x} -{\bf x}') {\cal \bf P}_{j}({\bf x}') = {\cal E}[{\bf P}],
\end{eqnarray}
and obviously pertains to the non-electrostatic structural component of the interactions, featuring explictly in Eq. (\ref{Xi-2}). Assuming the osmotic pressure of the form Eq. (\ref{Xi-2}), the two saddle-point equations for the auxiliary fields can then be cast as
\begin{eqnarray}
 &&{-{\cal \bf P}} =  {\lambda p^2}~  \frac{1}{u}~\frac{d}{du} \Big(\frac{\sinh{u}}{u}\Big) \left( {{\cal E}[{\bf P}]-\bnabla \phi}\right) \nonumber\\
&&\bnabla \Big(\varepsilon_0 \varepsilon_{\infty} \bnabla \phi  +  {\cal \bf P}\Big)  = 2\beta e \lambda_s \sinh{\beta e_0 \phi},
 \label{saddlerho}
\end{eqnarray}
with $u = \vert {{\cal E}[{\bf P}]-\bnabla \phi}\vert$. Clearly the first equation above plays the role of the first equation of Eq. (\ref{BjerrumPB}), but in this case it amounts to a non-linear and non-local constitutive equation. In the case of purely electrostatic coupling, {\sl i.e.} ${\cal E} = 0$, the above equation reduces to the Langevin-Poisson-Boltzmann equation that corresponds to a Coulomb gas in a dipolar solvent \cite{Abrashkin2007} and was investigated intensively in different contexts  \cite{Buyukdagli2013,Buyukdagli2014,Frydel2016,Buyukdagli2016, Budkov2018}.

After linearizing the above equations and introducing the  polarization potential $\phi_c$ defined in Eq. (\ref{potentials}), and then approximating the structural dipolar interaction as
\begin{eqnarray}
\tilde u_{ij}({\bf x} -{\bf x}') = \tilde u_P(0) \Big( \delta_{ij}  + \widehat{\xi}^2~\bnabla'_j\bnabla_i  \Big)\delta({\bf x} -{\bf x}') +  \dots \nonumber
\label{shortrange1}
\end{eqnarray}
where $\tilde u_P(0), \widehat{\xi}$ are the model constants, so that instead of the Slater-Takagi-Bethe form of the dielectric function of ice, we end up with
\begin{equation}
     \varepsilon_0 (\varepsilon  - \varepsilon_{\infty}) =  \frac{{\textstyle\frac{1}{3}} \lambda p^2}{1 + u_P(0){\textstyle\frac{1}{3}} \lambda p^2}, 
\label{def-epsilon2}     
\end{equation}
allowing us to write
\begin{eqnarray}
{\bf P} = \varepsilon_0(\varepsilon - \varepsilon_{\infty}) \bnabla \phi_c.
\end{eqnarray}
Note the difference in the definition of $\phi_c$ above and in Eq. (\ref{potentials}), where the polarization potential is connected only with the configurational part of polarization. We are then lead to a system of two linearized equations \cite{Blossey2022} for two independent degrees of freedom 
\begin{eqnarray}
\bnabla^2\phi_c  &=& {{\xi}^{-2}} \left( \phi_c - \phi\right),\\
 \bnabla^2\phi   &=& ~ {\textstyle\frac{\varepsilon}{\varepsilon_{\infty}}} \lambda_D^{-2} \phi +\,\, \left( {\textstyle\frac{\varepsilon}{\varepsilon_{\infty}}} - 1 \right) {{\xi}^{-2}} \left( \phi  - \phi_c \right)\, , \nonumber
\label{sp3c}
\end{eqnarray}
in which the inverse square of Debye length is defined by $\lambda_D^{-2} \equiv {2(\beta e)^2 \lambda_s}/{\varepsilon\varepsilon_0}$ and the structural correlation length is defined as ${\xi}^2 = (\varepsilon - \varepsilon_{\infty})\varepsilon_0 u_P(0)\, \widehat{\xi}^2$. Apart from the inconsequential difference in the definition of $\phi_c$, these equations are completely equivalent to the Onsager-Dupuis equations \cite{Onsager1960} and the same is true for the characteristic exponents as well as the non-local dielectric function.

The foregoing analysis is straightforwardly generalizable to more complex (higher order and/or non-quadratic) structural dipolar interactions as well as to the effects of ion hydration, {\sl i.e.} bound dipolar shell surrounding the ions \cite{Ben-Yaakov2011,Blossey2022}. 

\section{Conclusions and Outlook} 

Based on the microscopic theory of the dielectric properties of ice and with the guidance of the continuum phenomenological theories of non-local dielectrics, one can construct a field theory that can describe the non-locality and non-linearity of the dielectric response in aqueous media. The components of this theory are the electrostatic interactions between charges and the structural interactions between dipoles. The statistical mechanical derivation then yields two coupled mean-field (saddle-point) equations: a non-local and non-linear constitutive equations connecting the polarization and the electric field vectors, and the generalized  Poisson-Boltzmann equation connecting the polarization vector with the electrostatic potential.

These two equations clearly embody the fact that the polarization and the electric field vectors correspond to two independent degrees of freedom. In the case of no structural interactions, they reduce and combine into a single Langevin-Poisson-Boltzmann equation that was derived before in the case of pure electrostatic interactions, but represent a very different physics in the general case. In some sense they are closely related to the two equations for the polarization and electrostatic potentials in the Onsager-Dupuis theory of ice. The existence of these two independent degrees of freedom has also important repercussions in terms of the proper boundary conditions which we relegate to a future publication \cite{Blossey2022-2}. We also note that the steric constraints are easily implementable in our formalism through the lattice gas model just as in the case of an electrolyte \cite{Borukhov1997} or an electrolyte in a dipolar solvent \cite{Abrashkin2007}. This too will be discussed in more detail in a future publication \cite{Blossey2022-2}.

We expect that our non-local, non-linear theory of structured dielectrics as reviewed above should be particularly relevant and useful in the context of experiments on surface and interface interactions. Surface force apparatus (SFA), atomic force microscopy (AFM) or the colloidal probe technique 
based on AFM have recently demonstrated their value in probing the balance of forces and their interdependence between short range hydration structural oscillatory forces and longer ranged  Derjaguin-Landau-Verwey-Overbeek forces at solid-liquid interfaces \cite{Ludwig2020, vanLin2019, Scaratt2021,Klaassen2022,Kumar2022}. However, the measured interaction contains many unrelated as well as related effects, stemming in general from non-electrostatic couplings in the system\footnote{In some sense all interactions, even what we refer to as non-electrostatic interactions, are in fact electrostatic (plus quantum mechanics) in nature.} 
such as the chemical nature of ions, their size and charge, as well as polarizability, and solvent structuring hydrogen bonds, to invoke just a few \cite{Klaassen2022}.
It seems to us that the combination of experiments and a more systematic and general theory will allow significant further advances in a quantitative understanding of the
complexities of liquids at surfaces. In this sense analyzing theories in which these various non-electrostatic effects can be systematically decoupled into various well defined components of the total, measurable interaction, seems like a worthwhile direction to follow.
\\

{\bf Acknowledgements.}
\\

R.P. acknowledges the support of the University of Chinese Academy of Sciences and funding from the NSFC under Grant No. 12034019.

%\bibliographystyle{eplbib}
%\bibliography{structured-dielectrics}

\end{document}